\documentclass[10pt,oneside,final]{IEEEtran}

\usepackage{amssymb}
\usepackage{amsmath}

\usepackage{graphicx}
\usepackage{color}
\usepackage{multicol}

\usepackage{subfigure}
\usepackage{bm}
\usepackage{latexsym}
\usepackage{stfloats}
\usepackage{float}
\usepackage{exscale}
\usepackage{relsize}
\usepackage{cite}
\usepackage{cases}
\usepackage{algorithm}
\usepackage{algpseudocode}

\usepackage{epsfig}
\usepackage{epstopdf}
\usepackage{breqn}
\usepackage{enumerate}
\usepackage{multirow}
\usepackage[utf8]{inputenc}

\usepackage{mathtools}

\usepackage{algorithmicx}
\usepackage{setspace}
\usepackage{url}
\usepackage{makecell}

\pdfoutput=1

\begin{document}

\title{\vspace{-1.6em}Active Reconfigurable Intelligent Surface for Mobile Edge Computing
\vspace{-0.2em}}
\author{
	Zhangjie~Peng,
	Ruisong~Weng,
	Zhenkun~Zhang,
	\\
	Cunhua Pan,~\IEEEmembership{Member,~IEEE},
	and Jiangzhou~Wang,~\IEEEmembership{Fellow,~IEEE}
	\vspace{-0.5cm}
	\thanks{Z. Peng, R. Weng and Z. Zhang are with the College of Information, Mechanical and Electrical Engineering,
		Shanghai Normal University, Shanghai 200234, China (e-mails: pengzhangjie@shnu.edu.cn; 1000497102@smail.shnu.edu.cn;  1000479070@smail.shnu.edu.cn).}
	\thanks{C. Pan is with the National Mobile Communications Research Laboratory, Southeast University, China. (e-mail: cpan@seu.edu.cn).}
	\thanks{J. Wang is with the School of Engineering, University of Kent, CT2 7NT Canterbury, U.K. (e-mail: j.z.wang@kent.ac.uk).}
	\vspace{-0.8cm}
}
%

%
%
\maketitle

\newtheorem{lemma}{Lemma}
\newtheorem{theorem}{Theorem}
\newtheorem{remark}{Remark}
\newtheorem{corollary}{Corollary}
\newtheorem{proposition}{Proposition}

%
\vspace{-0.5cm}
\begin{abstract}
This paper investigates an active reconfigurable intelligent surface (RIS)-aided mobile edge computing (MEC) system. 
Compared with passive RIS, the active RIS is equipped with active reflective amplifier, which can effectively circumvent the ``double path loss" attenuation.
We propose a joint computing and communication design to minimize the maximum computational latency (MCL), subject to both the phase shift constraints and the edge computing capability constraints.
Specifically, the original problem is decoupled into four subproblems, and then the block coordinate descent (BCD) method and the successive convex approximation (SCA) method are applied to alternately optimize the subproblems.
The simulation results show that with the same power budget, the performance gain achieved by the active RIS is much larger than that by the passive RIS.


\end{abstract}

%

%

%
\vspace{-0.7cm}
\section{Introduction}
With the rapid increase in the amount of data generated by the Internet of Things (IoT), the traditional cloud computing is difficult to meet the high quality of service (QoS) requirements of the emerging applications \cite{7488250}.
Mobile edge computing (MEC), a new paradigm that is regarded as one of the most promising techniques in the future communication system, was proposed in \cite{8016573}.
In MEC systems, powerful and reliable computing resources are deployed at the edge of the wireless networks to serve the mobile devices, enabling them to execute computation-intensive and latency-critical applications.
However, the computation offloading links may suffer from severe signal attenuation or blockage, which will increase the offloading latency, and inevitably deteriorate the overall performance of MEC system \cite{9690150}.

Recently, reconfigurable intelligent surfaces (RISs) has been envisioned as an effective solution for enhancing the channel gains of offloading links in MEC systems \cite{9326394}.
RIS is a thin planar board that is composed of an array of low-cost and passive  reflecting elements integrated with controllable electronics.
By reconfiguring the electromagnetic propagation environment and constructing the reflection links, RISs can effectively improve the offloading data rate of the edge devices in MEC systems \cite{9133107,9270605}.
However, when an RIS is deployed between the receiver and the transmitter with favourable communication link, the performance gain is limited \cite{zhang2021active}.
Besides, the potential of RIS is significantly restricted by the inevitable ``double path loss" attenuation in the reflection link, i.e., the signal received through this link will suffer from the large-scale fading twice \cite{9734027}.
To this end, the concept of active RIS, which is expected to circumvent these disadvantages, has been proposed in \cite{9734027,https://doi.org/10.48550/arxiv.2106.10963,https://doi.org/10.48550/arxiv.2201.12565,9840889}.
Different from passive RIS, the active RIS is equipped with an integrated active reflection-type amplifier, which is able to amplify the received signal by consuming additional power.
The application of this device has been explored in various scenarios, such as multiple-input single-output (MISO) systems\cite{9734027}, multiple-input multiple-output (MIMO) systems\cite{zhang2021active} and IoT networks \cite{https://doi.org/10.48550/arxiv.2201.12565}, and the results in [8]-[10] showed that active RIS can achieve  significant performance gain in terms of data rate \cite{zhang2021active,9377648,https://doi.org/10.48550/arxiv.2106.10963} and physical layer security \cite{9652031,9530403}.

However, there is a lack of studies on the use of active RIS in MEC systems. 
In this paper, we propose to minimize the maximum computational latency (MCL) of the MEC system by jointly optimizing the receive beamforming of the access point (AP), the reflection coefficients of the active RIS, the computation offloading volume and the computing resource allocation.
The MCL minimization problem is solved by applying the block coordinate descent (BCD) method and the successive convex approximation (SCA) method. 
The simulation results not only verify the effectiveness of our algorithm, but also show that the performance gain achieved by the active RIS is much larger than that by using the passive RIS with the same power budget.

\vspace{-0.5cm}
\section{System Model}

\begin{figure}
	\vspace{0.1cm}
	\centering
	\includegraphics[width=0.8\linewidth]{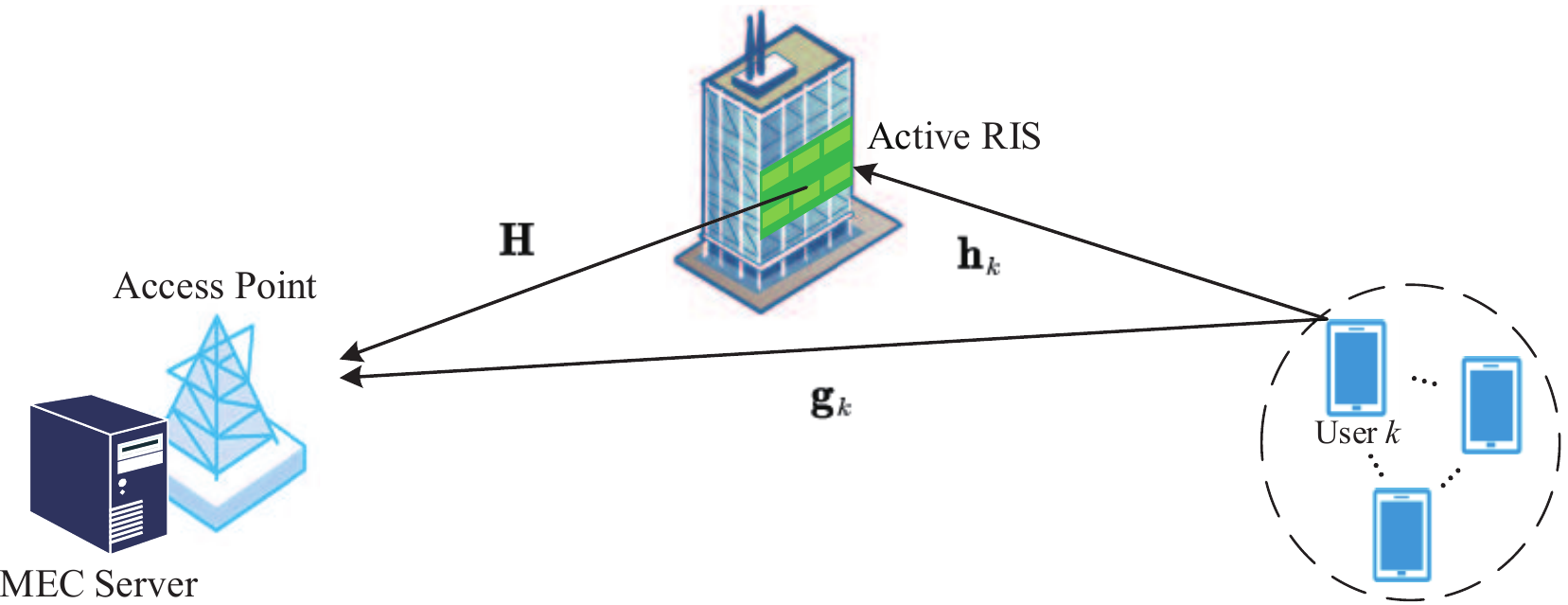}
	\caption{An active RIS-aided MEC system with an $N$-antenna AP and $K$ single-antenna users.}
	\label{system_model}
	\vspace{-0.8cm}
\end{figure}

\subsection{Signal Transmission Model}
Consider an MEC system as shown in Fig. \ref{system_model}, which consists of an AP equipped with $N$ receive antennas and $K$ single-antenna users.
An MEC server has a wired connection to the AP with negligible data transmission latency, to which each user can offload a certain fraction of or all of its computational tasks through the communication link.
To assist the users’ computation offloading, an active RIS with $M$ reflecting elements is deployed to reconfigure the propagation environment and enhance the reflected signals. 

The baseband equivalent channels from the RIS to the AP, from the $k$-th user to the RIS, and from the $k$-th user to the AP are denoted by $\mathbf{H}\in \mathbb{C}^{N\times M}$, $\mathbf{h}_k\in \mathbb{C}^{M\times 1}$ and $\mathbf{g}_k\in \mathbb{C}^{N\times 1}$, respectively.
$\theta _m=\lambda _me^{j\phi _m}$ is the reflection coefficient of the $m$-th reflecting element of the RIS, and the phase shift $\phi _m\in \left[ 0,2\pi \right] $.
Then, the reflecting beamforming of the RIS can be represented as the reflection coefficient vector $\boldsymbol{\theta }=\left[ \theta _1,\theta _2,\dotsc,\theta _M \right] ^{\mathrm{T}}$.
Different from passive RISs, the reflecting elements of an active RIS are equipped with amplifiers that consume additional power, and thus the thermal noise at the active RIS is not negligible \cite{zhang2021active}.
Hence, the signal received at the AP is modelled as 
\setlength\abovedisplayskip{0.3pt}
\setlength\belowdisplayskip{0.3pt}
\begin{equation}
	\mathbf{y}=\sum_{k=1}^K{\sqrt{p_k}\left( \mathbf{H\Theta h}_k+\mathbf{g}_k \right) s_k+\mathbf{H\Theta v}+\mathbf{n}},
\end{equation}
where $\mathbf{\Theta }\triangleq \mathrm{diag}\left( \boldsymbol{\theta } \right)$, $s_k$ is the data symbol sent by the $k$-th user, $p_k$ denotes the corresponding computation offloading power, and $\mathbf{v}\in \mathbb{C} ^{M\times 1}\sim \mathcal{C}\mathcal{N}\left( \mathbf{0}_M,\sigma ^2\mathbf{I}_M \right)$ and $\mathbf{n}\in \mathbb{C} ^{N\times 1}\sim\mathcal{C}\mathcal{N}\left( \mathbf{0}_N,\delta ^2\mathbf{I}_N \right)$ denote the thermal noise at the RIS and the AP, respectively.
Define $\mathcal{M} \!=\! \left\{ 1,2,\dotsc,M \right\}$ and $\mathcal{K} = \left\{ 1,2,\dotsc,K \right\}$.
It is assumed that each $s_k\sim \mathcal{C}\mathcal{N}\left( 0,1 \right) 
$ for $\forall k \in \mathcal{K}$.

We assume that the AP decodes the received signal by using linear receive beamforming.
By denoting $\mathbf{F}=\left[ \mathbf{f}_1,\dotsc,\mathbf{f}_K \right] \in \mathbb{C}^{N\times K}$ as the set of the receiving beamforming vectors, the signal recovered at the AP is given by
\begin{equation}\label{s_hat}
	\mathbf{\hat{s}}=\mathbf{F}^{\mathrm{H}}\mathbf{y},
\end{equation}
where $\mathbf{\hat{s}}=\left[ \hat{s}_1,\dotsc,\hat{s}_K \right] ^{\mathrm{T}}$ and $\hat{s}_k$ is the recovered signal for the $k$-th user.
Then, the achievable computation offloading rate (bit/s) of the $k$-th user is given by
\begin{equation}
	R_k=B\log_2 \left( 1+\gamma _k \right),
\end{equation}
where $B$ is the channel bandwidth and
\begin{equation}
	\gamma _k=\frac{p_k\left| \mathbf{f}_{k}^{\mathrm{H}}\left( \mathbf{H\Theta h}_k+\mathbf{g}_k \right) \right|^2}{\sum\limits_{\substack{i=1 \\ i\ne k}}^K{p_i\left| \mathbf{f}_{k}^{\mathrm{H}}\left( \mathbf{H\Theta h}_i+\mathbf{g}_i \right) \right|^2}+\sigma^2\left\| \mathbf{f}_{k}^{\mathrm{H}}\mathbf{H\Theta } \right\| ^2+\delta ^2\left\| \mathbf{f}_{k}^{\mathrm{H}} \right\| ^2}.
\end{equation}

\vspace{-0.9cm}
\subsection{Computation Task Model}
We consider the partial offloading scheme by applying the data-partition model \cite{8016573}, which is widely used in the existing literature \cite{9133107,7542156}.
Specifically, the task-input bits are assumed to be bit-wise independent.
Each user divides its computational task into two parts, which are respectively handled locally and offloaded to the edge computing node.
The latency model of the considered MEC system is detailed as follows:

\subsubsection{Local Computing}
The total computational task and the offloading volume of the $k$-th user in terms of the number of bits are denoted by $L_k$ and $l_k$, respectively.
In addition, the computing power of the $k$-th user is quantified by its number of central processing unit (CPU) cycles per second $f_{\mathrm{L},k}$. 
Denote by $c_k$ the number of CPU cycles required to process each bit. 
Then, the latency at the $k$-th user induced by local computing can be expressed as
\vspace{-0.1cm}
\begin{equation}\label{T_Lk}
	T_{\mathrm{L},k}=\frac{\left( L_k-l_k \right) c_k}{f_{\mathrm{L},k}}.
\end{equation}

\subsubsection{Edge Computing}
The MEC server is assumed to start the process of the offloaded task of the $k$-th user at the time when all $l_k$ bits are completely received.
Compared with offloaded computational task, the amount of bits of the computational results is ususally considered to be much smaller, so the feedback delay can be reduced to a negligible level \cite{8472907}. 
Hence, the time required for executing the $k$-th user's offloaded computational task is mainly composed of that required for computation offloading and edge computing, i.e.,
\begin{equation}\label{T_ek}
	T_{\mathrm{E},k}=\frac{l_k}{R_k}+\frac{l_kc_k}{f_{\mathrm{E},k}},
\end{equation}
where $f_{\mathrm{E},k}$ (cycle/s) is the edge computing resource allocated to the $k$-th user.

It is assumed that the users are able to carry out the local computing and  computation offloading in parallel.
In this case,
the total latency of the $k$-th user can be formulated as
\begin{equation}
	T_k=\max \left\{ T_{\mathrm{L},k},T_{\mathrm{E},k} \right\}.
\end{equation}

\vspace{-0.4cm}
\subsection{Problem Formulation}
To solve the problem of minimizing the MCL of the MEC system, we propose to jointly optimize the receive beamforming matrix $\mathbf{F}$, the reflection coefficient vector $\boldsymbol{\theta }$, the computation offloading power $\mathbf{p}=\left[ \sqrt{p_1},\dotsc,\sqrt{p_K} \right]^{\mathrm{T}}$, the computation offloading volume $\boldsymbol{l}=\left[ l_1,\dotsc,l_K \right] ^{\mathrm{T}}$ and the computing resource allocation $\boldsymbol{f}_{\mathrm{E}}=\left[ f_{\mathrm{E},1},\dotsc,f_{\mathrm{E},K} \right] ^{\mathrm{T}}$.
Specifically, the MCL minimization problem is formulated as
\vspace{-0.5cm}
\begin{subequations}\label{MCL_problem}
	\begin{alignat}{2}
		\min_{\mathbf{F},\boldsymbol{\theta },\mathbf{p},\boldsymbol{l},\boldsymbol{f}_{\mathrm{E}}} \quad& \max_{k\in \mathcal{K}}\left\lbrace  T_k\right\rbrace 
		\\
		\mbox{s.t.}\quad\;\;\;
		& p_k\in \left[0,p_{max}\right], k\in\mathcal{K}, \label{constraint_p}
		\\
		& l_k\in \left\{ 0,1,\dotsc,L_k \right\}, k\in\mathcal{K}, \label{constraint_l_k}
		\\
		& f_{\mathrm{E},k}\geqslant 0,\;k\in\mathcal{K},\label{constraint_f_E>0}
		\\
		& \sum_{k=1}^K{f_{\mathrm{E},k}}\leqslant f_{\mathrm{E}}^{\mathrm{tot}},\label{constraint_f_E}
		\\
		& \sum_{k=1}^K{p_k\left\| \mathbf{\Theta h}_k \right\| ^2}+\left\| \mathbf{\Theta } \right\| ^2\sigma ^2\leqslant P_{\mathrm{aRIS}},\label{constraint_P}
	\end{alignat}
\end{subequations}
where $f_{\mathrm{E}}^{\mathrm{tot}}$ denote the computing power of the MEC server.
In the power constraint \eqref{constraint_P}, the amplification power budget is modelled as $P_{\mathrm{aRIS}}=\xi \left( P_{\mathrm{tot}}-M\left( P_{\mathrm{DC}}+P_{\mathrm{c}} \right) \right) $, where $\xi$, $P_{\mathrm{tot}}$, $P_{\mathrm{DC}}$ and $P_{\mathrm{c}}$ are the amplifier efficiency, the total power budget, the DC biasing power consumption and the circuit power consumption of the active RIS, respectively \cite{9377648}.

\vspace{-0.3cm}
\section{Joint Computing and Communication Design}
In this section, an effective algorithm is proposed to solve the MCL minimization problem \eqref{MCL_problem} based on the BCD method,
where the variables are optimized alternately.

\vspace{-0.5cm}
\subsection{Local and Edge Computing Design}\label{Local and Edge Computing Design}
By fixing the receive beamforming matrix $\mathbf{F}$, the reflection coefficient vector $\boldsymbol{\theta }$ and the computation offloading power $\mathbf{p}$, we can formulate the subproblem for local and edge computing design as follows
\vspace{-0.2cm}
\begin{alignat}{2}
	\min_{\boldsymbol{l},\boldsymbol{f}_{\mathrm{E}}} \quad& \max_{k\in \mathcal{K}}\left\lbrace  T_k\right\rbrace \label{MCL_problem_0}
	\\
	\mbox{s.t.}\quad&\eqref{constraint_l_k} ,\eqref{constraint_f_E>0},\eqref{constraint_f_E}.\notag
\end{alignat}
In the following, Problem \eqref{MCL_problem_0} is solved by alternately optimizing each subset of the variables.

\subsubsection{Optimizing the computation offloading volume $\boldsymbol{l}$}
Note that the feasible region of $\boldsymbol{l}$ is discrete.
By relaxing $\boldsymbol{l}$ to continuous variables, i.e.,  $\tilde{l}_k\in \left[ 0,L_k \right]$ and $\boldsymbol{\tilde{l}}=\left[ \tilde{l}_1,\dotsc,\tilde{l}_K \right] ^{\mathrm{T}}$, we have
\begin{align}
	&T_k\left( \tilde{l}_k \right) =\max \left\{ T_{\mathrm{L},k}(\tilde{l}_k),T_{\mathrm{E},k}(\tilde{l}_k) \right\} \notag
	\\
	&=\begin{cases}
		\frac{\left( L_k-\tilde{l}_k \right) c_k}{f_{\mathrm{L},k}},\!\!&		0\leqslant \tilde{l}_k\leqslant \frac{L_kc_kR_kf_{\mathrm{E},k}}{f_{\mathrm{L},k}f_{\mathrm{E},k}+c_kR_k\left( f_{\mathrm{L},k}+f_{\mathrm{E},k} \right)}\\
		\frac{\tilde{l}_k}{R_k}+\frac{\tilde{l}_kc_k}{f_{\mathrm{E},k}},\!\!&		\frac{L_kc_kR_kf_{\mathrm{E},k}}{f_{\mathrm{L},k}f_{\mathrm{E},k}+c_kR_k\left( f_{\mathrm{L},k}+f_{\mathrm{E},k} \right)}\!\leqslant\! \tilde{l}_k\leqslant L_k\\
	\end{cases}\!.
\end{align}
It can be readily verified that $T_{k}\left( \tilde{l}_k \right) $ is minimized at
\begin{equation}\label{l_tilde}
	\tilde{l}_k^{\mathrm{opt}} = \frac{L_kc_kR_kf_{\mathrm{E},k}}{f_{\mathrm{L},k}f_{\mathrm{E},k}+c_kR_k\left( f_{\mathrm{L},k}+f_{\mathrm{E},k} \right)}.
\end{equation}
Then, the optimal $l_k$ can be obtained as
\begin{equation}\label{l_opt}
	l_{k}^{\mathrm{opt}}={\mathrm{arg}\min}_{\tilde{l}_k\in \left\{ \lfloor \tilde{l}_{k}^{\mathrm{opt}} \rfloor ,\lceil \tilde{l}_{k}^{\mathrm{opt}} \rceil \right\}}\,\,T_{k}\left( \tilde{l}_k \right) ,
\end{equation}
where $\lceil \cdot \rceil$ and $\lfloor \cdot \rfloor$ represent the ceiling and floor operations, respectively.

\vspace{-0.1cm}
\subsubsection{Optimizing the computing resource allocation $\boldsymbol{f}_{\mathrm{E}}$}
By letting $l_k=\tilde{l}_{k}^{\mathrm{opt}}$, $k\in\mathcal{K}$ in Problem \eqref{MCL_problem_0}, the subproblem corresponding to $\boldsymbol{f}_{\mathrm{E}}$ can be reformulated as 
\begin{alignat}{2}
	\min_{\boldsymbol{f}_{\mathrm{E}}} \quad& \max_{k\in \mathcal{K}} \left\lbrace  \frac{L_kc_kf_{\mathrm{E},k}+L_kc_{k}^{2}R_k}{f_{\mathrm{L},k}f_{\mathrm{E},k}+c_kR_k\left( f_{\mathrm{L},k}+f_{\mathrm{E},k} \right)}\right\rbrace \label{MCL_problem_f}
	\\ 
	\mbox{s.t.}\quad&\eqref{constraint_f_E>0},\eqref{constraint_f_E}.\notag
\end{alignat}
By introducing an auxiliary variable $\eta$, Problem \eqref{MCL_problem_f} can be transformed as
\begin{subequations}\label{MCL_problem_f1}
	\begin{align}
		\min_{\eta ,\boldsymbol{f}_{\mathrm{E}}} \quad& \eta 
		\\
		\mbox{s.t.}\quad& \frac{L_kc_kf_{\mathrm{E},k}+L_kc_{k}^{2}R_k}{f_{\mathrm{L},k}f_{\mathrm{E},k}+c_kR_k\left( f_{\mathrm{L},k}+f_{\mathrm{E},k} \right)}\leqslant \eta, \; \forall k\in \mathcal{K},\label{constraint_T<eta}
		\\
		&\eqref{constraint_f_E>0},\eqref{constraint_f_E}.\notag
	\end{align}
\end{subequations}
To solve this problem, the method based on the SCA method is applied.
Firstly, by introducing auxiliary variables $\boldsymbol{a}=\left[ a_1,\dotsc,a_K \right] ^{\mathrm{T}}
$ and $\boldsymbol{b}=\left[ b_1,\dotsc,b_K \right] ^{\mathrm{T}}
$, constraint \eqref{constraint_T<eta} can be replaced with the following constraints
\begin{align}
&\frac{L_kc_k}{a_k}\leqslant \frac{c_kR_kf_{\mathrm{L},k}}{f_{\mathrm{E},k}}+f_{\mathrm{L},k}+c_kR_k \!\triangleq\! g_{k}(f_{\mathrm{E},k}),\forall k\in \mathcal{K},\label{constraint_SCA_a_origin}
\\
&\frac{L_kc_{k}^{2}R_k}{b_k}\leqslant \left( f_{\mathrm{L},k}+c_kR_k \right) f_{\mathrm{E},k}+c_kR_kf_{\mathrm{L},k},\forall k\in \mathcal{K},\label{constraint_SCA_b}
\\
&a_k+b_k\leqslant \eta ,\;\forall k\in \mathcal{K}.\label{constraint_SCA_eta}
\end{align} 
Note that constraint \eqref{constraint_SCA_a_origin} is non-convex.
Then, by utilizing the inequality (3.2) in \cite{convexBook} that a convex function is lower-bounded by its first-order Taylor expansion, we have
\begin{align}
&\frac{L_kc_k}{a_k}\leqslant \frac{c_kR_kf_{\mathrm{L},k}}{f_{\mathrm{E},k}^{r}}-\frac{c_kR_kf_{\mathrm{L},k}}{f_{\mathrm{E},k}^{r,2}}\left( f_{\mathrm{E},k}-f_{\mathrm{E},k}^{r} \right) +f_{\mathrm{L},k}+c_kR_k \notag\label{constraint_SCA_a}
\\
&\triangleq h_k\left( f_{\mathrm{E},k} \mid f_{\mathrm{E},k}^{r} \right), \forall k\in \mathcal{K},
\end{align}
where $f_{\mathrm{E},k}^{r}$ is the value of $f_{\mathrm{E},k}$ at the $r$-th iteration.
It can be readily verified that when $f_{\mathrm{E},k} = f_{\mathrm{E},k}^{r}$, $h_k( f_{\mathrm{E},k} \mid f_{\mathrm{E},k}^{r})=g_{k}(f_{\mathrm{E},k})$.
Finally, the problem of $\boldsymbol{f}_{\mathrm{E}}$ to be solved at the $r$-th SCA iteration is formulated as
\begin{align}
	\label{f_problem}\min_{\eta ,\boldsymbol{f}_{\mathrm{E}},\boldsymbol{a},\boldsymbol{b}} \quad& \eta 
	\\
	\mbox{s.t.}\quad\;\;\;& \eqref{constraint_f_E>0},\eqref{constraint_f_E} ,\eqref{constraint_SCA_b}-\eqref{constraint_SCA_a}. \notag
\end{align}
The globally optimal solution of this problem can be obtained by the common optimization tools, e.g., CVX.
By denoting the objective function of Problem \eqref{MCL_problem} as $\mathcal{T}\left(\mathbf{F},\boldsymbol{\theta },\boldsymbol{l},\boldsymbol{f}_{\mathrm{E}}\right) $,
the detail of the proposed SCA algorithm for solving Problem \eqref{f_problem} is summarized in Algorithm \ref{SCA}, which is guaranteed to  converge to the Karush-Kuhn-Tucker solution of Problem \eqref{MCL_problem_f1}, as proved in \cite{9123680}.

\begin{algorithm}[t]
	\caption{SCA Algorithm for solving Problem \eqref{f_problem}}
	\label{SCA}
	\textbf{Initialize}: Initialize  the number of iterations as $r=1$, set feasible $\boldsymbol{f}_{\mathrm{E}}^1$, the maximum number of iterations $r_{\max}$ and the error tolerance  $\varrho$.
	\begin{algorithmic}[1]
		\While{$\left| \mathcal{T}\left(\boldsymbol{f}_{\mathrm{E}}^{r+1} \right) -\mathcal{T}\left(\boldsymbol{f}_{\mathrm{E}}^{r} \right) \right|/\mathcal{T}\left( \boldsymbol{f}_{\mathrm{E}}^{r} \right) \geqslant \varrho$ and $n\leqslant r_{\max}$}
		\State Calculate $\boldsymbol{f}_{\mathrm{E}}^{r+1}$ by solving Problem \eqref{f_problem};
		\State Set $r \gets r+1$;
		\EndWhile
	\end{algorithmic}
\end{algorithm}

\subsection{Signal Transmission Design}
Given the computation offloading volume $\boldsymbol{l}$ and the computing resource allocation $\boldsymbol{f}_{\mathrm{E}}$, the optimal signal transmission design can be obtained by solving the following subproblem
	\begin{align}
		\min_{\mathbf{F},\boldsymbol{\theta },\mathbf{p}} \quad & \max_{k\in \mathcal{K}}\left\lbrace  T_k\right\rbrace \\ \label{MCL_problem_1}
		\mbox{s.t.}\quad&\eqref{constraint_p},\eqref{constraint_P} \notag
	\end{align}
As discussed in Section \ref{Local and Edge Computing Design}, the optimal value of the objective function of Problem \eqref{MCL_problem} satisfies $T_{k} ^{\mathrm{opt}} = T_{\mathrm{L},k} ^{\mathrm{opt}} = T_{\mathrm{E},k} ^{\mathrm{opt}}$.
In addition, the expression of $T_{\mathrm{L},k}$ in \eqref{T_Lk} is independent of $\mathbf{F}$ and $\boldsymbol{\theta }$.
Thus, we replace $T_{k}$ with $T_{\mathrm{E},k}$ in \eqref{MCL_problem_1}.
Furthermore, by introducing an auxiliary variable $\varepsilon$, Problem \eqref{MCL_problem_1} can be transformed as
\vspace{-0.2cm}
\begin{subequations}\label{MRM_problem}
	\begin{align}
		\min_{\mathbf{F},\boldsymbol{\theta },\mathbf{p},\varepsilon} \quad& \varepsilon 
		\\
		\mbox{s.t.}\quad\;
		&\frac{l_k}{R_k}+\frac{l_kc_k}{f_{\mathrm{E},k}}\leqslant\varepsilon,\;\forall k\in \mathcal{K},\label{constraint_TE>epsilon}
		\\
		& \eqref{constraint_p},\eqref{constraint_P}.\notag
	\end{align}
\end{subequations}
Note that constraint \eqref{constraint_P} is convex.
Next, we transform constraint \eqref{constraint_TE>epsilon} by utilizing the equivalence between data rate and the mean-square error \cite{5756489}.
From \eqref{s_hat}, we can derive the mean-square error of the signal recovered at the AP corresponding to the $k$-th user as follows 
\begin{align}\label{MSE_user}
	d_k&= \sum_{i=1}^K{\left|\sqrt{p_i}\mathbf{f}_{k}^{\mathrm{H}}\left( \mathbf{H\Theta h}_i+\mathbf{g}_i \right)\right|^2} + \sigma ^2 \left\| \mathbf{f}_{k}^{\mathrm{H}}\mathbf{H\Theta } \right\| ^2+ \delta ^2 \left\| \mathbf{f}_{k}^{\mathrm{H}} \right\| ^2\notag
	\\
	&\quad-2\mathrm{Re}\left\{ \sqrt{p_k}\mathbf{f}_{k}^{\mathrm{H}}\left( \mathbf{H\Theta h}_k+\mathbf{g}_k \right) \right\} +1. 
\end{align}
By introducing a set of auxiliary variables  $\mathcal{V}=\left\{ v_k\geqslant 0,k\in \mathcal{K} \right\} $, $R_k$ can be reformulated as
\begin{equation}\label{MSE_user_speed}
	\tilde{R}_k(\mathbf{F},\mathbf{\Theta },\mathbf{p},\mathcal{V})=B\left( \log _2\left| v_k \right|-\frac{v_kd_k}{\log _{\mathrm{e}}2}+\frac{1}{\log _{\mathrm{e}}2} \right).
\end{equation}
Note that $\tilde{R}_k(\mathbf{F},\mathbf{\Theta },\mathbf{p},\mathcal{V})$ is a concave function with respect to each variable when the others are fixed,
which is more tractable than $R_k$.
Therefore, Problem \eqref{MRM_problem} can be reformulated as
\vspace{-0.3cm}
\begin{subequations}\label{MRM_problem_1}
	\begin{align}
		\min_{\mathbf{F},\boldsymbol{\theta },\mathbf{p},\mathcal{V},\varepsilon} \quad& \varepsilon 
		\\
		\mbox{s.t.}\quad\;\;& \tilde{R}_k\geqslant \frac{l_k{f_{\mathrm{E},k}}}{\varepsilon{f_{\mathrm{E},k}} -l_kc_k},\;\forall k\in \mathcal{K},
		\\
		&\eqref{constraint_p},\eqref{constraint_P}.	\notag
	\end{align}
\end{subequations}
Then, we alternately optimize the objective function over $\mathcal{V}$, $\mathbf{F}$, $\boldsymbol{\theta }$ and $\mathbf{p}$ to solve Problem \eqref{MRM_problem_1}.

\subsubsection{Optimizing the Auxiliary Variables  $\mathcal{V}$ and the Receive Beamforming Matrix $\mathbf{F}$}
For $k\in \mathcal{K}$, the optimal $v_k$ can be obtained by setting the first-order derivative of $\tilde{R}_k(\mathbf{F},\boldsymbol{\lambda },\boldsymbol{\phi},\mathcal{V})$ with respect to $v_k$ to zero, which is given by
\begin{align}
	v_{k} ^{\mathrm{opt}} = d_{k} ^{-1}.
\end{align}
Similarly, when the other variables are fixed, the optimal $\mathbf{f}_k$ can be derived as
\begin{align}
	\mathbf{f}_{k}^{\mathrm{opt}} &= \sqrt{p_k}\left(  \sum_{i=1}^K{p_i \left(  \mathbf{H\Lambda \Phi h}_i + \mathbf{g}_i  \right)  \left( \mathbf{H\Lambda \Phi h}_i + \mathbf{g}_i  \right)  ^{\mathrm{H}}} + \delta ^2\mathbf{I}_N \right. \notag
	\\
	&\quad+\left.  \sigma ^2\mathbf{H\Lambda \Phi \Phi }^{\mathrm{H}}\mathbf{\Lambda }^{\mathrm{H}}\mathbf{H}^{\mathrm{H}} \right) ^{-1} \left( \mathbf{H\Lambda \Phi h}_k+\mathbf{g}_k \right).
\end{align}

\subsubsection{Optimizing the Reflection Coefficient $\boldsymbol{\theta }$}
Based on the equality in \cite[Eq. (1.10.6)]{Zhang2017Matrix}, we can rewrite \eqref{MSE_user} as

\begin{align}\label{MSE_lambda}
	d_k&=\mathrm{Tr}\left( \mathbf{\Theta }^{\mathrm{H}}\mathbf{H}^{\mathrm{H}}\mathbf{f}_k\mathbf{f}_{k}^{\mathrm{H}}\mathbf{H\Theta A} \right) +\sigma ^2\mathrm{Tr}\left( \mathbf{\Theta }^{\mathrm{H}}\mathbf{H}^{\mathrm{H}}\mathbf{f}_k\mathbf{f}_{k}^{\mathrm{H}}\mathbf{H\Theta } \right) \notag
	\\
	&\quad+2\mathrm{Re}\left\{ \mathrm{Tr}\left( \mathbf{W}_k\mathbf{\Theta } \right) \right\} +a_k \notag
	\\
	&=\boldsymbol{\theta }^{\mathrm{H}}\mathbf{B}_k\boldsymbol{\theta }+2\mathrm{Re}\left\{ \mathbf{w}_k\boldsymbol{\theta } \right\} +a_k,
\end{align}
where
\begin{align*}
	\mathbf{A}&\triangleq \sum_{i=1}^K{p_i\mathbf{h}_i\mathbf{h}_i^{\mathrm{H}}},	\mathbf{w}_k\triangleq \left[ \left[ \mathbf{W}_k \right] _{1,1},\dotsc,\left[ \mathbf{W}_k \right] _{M,M} \right] ^{\mathrm{T}},
	\\
	\mathbf{W}_k&\triangleq \sum_{i=1}^K{p_i\mathbf{h}_i\mathbf{g}_i^{\mathrm{H}}}\mathbf{f}_k\mathbf{f}_{k}^{\mathrm{H}}\mathbf{H}-\sqrt{p_k}\mathbf{h}_k\mathbf{f}_{k}^{\mathrm{H}}\mathbf{H},
	\\
	a_k&\triangleq \mathbf{f}_{k}^{\mathrm{H}}\left( \sum_{i=1}^K{p_i\mathbf{g}_i\mathbf{g}_i^{\mathrm{H}}}\right) \mathbf{f}_k+\delta ^2\mathbf{f}_{k}^{\mathrm{H}}\mathbf{f}_k\!-\!2\mathrm{Re}\left\{ \sqrt{p_k}\mathbf{f}_{k}^{\mathrm{H}}\mathbf{g}_k \right\}\!+\!1,
	\\
	\mathbf{B}_k&\triangleq \left( \mathbf{H}^{\mathrm{H}}\mathbf{f}_k\mathbf{f}_{k}^{\mathrm{H}}\mathbf{H} \right) \odot \left( \mathbf{A}+\sigma ^2\mathbf{I}_M \right) ^{\mathrm{T}}.
\end{align*}

By substituting \eqref{MSE_lambda} into \eqref{MSE_user_speed}, we have
	\begin{equation}
		\bar{R}_k(\boldsymbol{\theta })= -\boldsymbol{\theta }^{\mathrm{H}}\mathbf{\bar{B}}_k\boldsymbol{\theta }-2\mathrm{Re}\left\{ \mathbf{\bar{w}}_{k}^{\mathrm{H}}\boldsymbol{\theta } \right\} +\bar{a}_k,
	\end{equation}
	where $\bar{a}_k\triangleq B\left( \log _2\left| v_k \right|-\frac{v_ka_k}{\log _{\mathrm{e}}2}+\frac{1}{\log _{\mathrm{e}}2} \right)$, $\mathbf{\bar{B}}_k\triangleq \frac{v_kB\mathbf{B}_k}{\log _{\mathrm{e}}2}$ and $\mathbf{\bar{w}}_k\triangleq \frac{v_kB\mathbf{w}_k}{\log _{\mathrm{e}}2}$.
	Similarly, constraint \eqref{constraint_P} can be rewritten as 
	\begin{equation}\label{constraint_theta}
		\boldsymbol{\theta }^{\mathrm{H}}\left( \mathbf{I}_M\odot \mathbf{A}^{\mathrm{T}}+\sigma ^2\mathbf{I}_M \right) \boldsymbol{\theta }\leqslant P_{\mathrm{aRIS}}.
	\end{equation}
	Then, the subproblem of $\boldsymbol{\theta}$ is formulated as
	\begin{subequations}\label{MRM_problem_2}
		\begin{align}
			\min_{\boldsymbol{\theta},\varepsilon} \quad& \varepsilon 
			\\
			\mbox{s.t.}\quad& \bar{R}_k\left( \boldsymbol{\theta} \right)\geqslant \frac{l_k{f_{\mathrm{E},k}}}{\varepsilon{f_{\mathrm{E},k}} -l_kc_k},\;\forall k\in \mathcal{K},
			\\
			&\eqref{constraint_theta}.\notag
		\end{align}
\end{subequations}
From \eqref{T_ek}, we have $\varepsilon\geqslant T_{\mathrm{E},k} >\frac{l_kc_k}{f_{\mathrm{E},k}}$.
Problem \eqref{MRM_problem_2} is a second-order cone programming problem, and its globally optimal solution can also be obtained by existing optimization tools.
\subsubsection{Optimizing the Computation Offloading Power $\mathbf{p}$} 
	We introduce a selection vector $\mathbf{t}_{k} \in \mathbb{R}^{K \times 1}$, in which the $k$-th element is equal to 1, while other elements are 0. Then, \eqref{MSE_user} can be rewritten as
	\begin{align}\label{MSE_user_p}
		d_k=\mathbf{p}^{\mathrm{T}}\mathbf{C}_k\mathbf{p}-2\mathrm{Re}\left\{ \mathbf{p}^{\mathrm{T}}\mathbf{j}_k \right\}+m_k,
	\end{align}
	where
	\vspace{-0.2cm}
	\begin{align*}
		\mathbf{C}_k&\triangleq\sum_{i=1}^K{\mathbf{t}_{k}\mathbf{f}_{k}^{\mathrm{H}}\left( \mathbf{H\Theta h}_i+\mathbf{g}_i \right)\left( \mathbf{H\Theta h}_i+\mathbf{g}_i \right)^{\mathrm{H}}\mathbf{f}_{k}\mathbf{t}_{k}^{\mathrm{H}}},
		\\
		\mathbf{j}_k&\triangleq\mathbf{t}_{k}\mathbf{f}_{k}^{\mathrm{H}}\left( \mathbf{H\Theta h}_k+\mathbf{g}_k \right),
		\\
		m_k&\triangleq\sigma ^2 \left\| \mathbf{f}_{k}^{\mathrm{H}}\mathbf{H\Theta } \right\| ^2+ \delta ^2 \left\| \mathbf{f}_{k}^{\mathrm{H}} \right\| ^2+1.
	\end{align*}
	By substituting \eqref{MSE_user_p} into \eqref{MSE_user_speed}, we have
	\begin{equation}
		\widehat{R}_k(\mathbf{p})= -\mathbf{p}^{\mathrm{T}}\mathbf{\widehat{C}}_k\mathbf{p}+2\mathrm{Re}\left\{ \mathbf{p}^{\mathrm{T}}\mathbf{\widehat{j}}_k \right\} +\widehat{m}_k,
	\end{equation}
	where $\widehat{m}_k\triangleq B\left( \log _2\left| v_k \right|-\frac{v_km_k}{\log _{\mathrm{e}}2}+\frac{1}{\log _{\mathrm{e}}2} \right)$, $\mathbf{\widehat{C}}_k\triangleq \frac{v_kB\mathbf{C}_k}{\log _{\mathrm{e}}2}$ and $\mathbf{\widehat{w}}_k\triangleq \frac{v_kB\mathbf{j}_k}{\log _{\mathrm{e}}2}$.
	
	Then, constraint \eqref{constraint_P} can be rewritten as 
	\begin{align}\label{constraint_P_p}
		\mathbf{p}^{\mathrm{T}}(\sum_{k=1}^K{\mathbf{t}_{k}\left\| \mathbf{\Theta h}_k \right\| ^2}\mathbf{t}_{k}^{\mathrm{H}})\mathbf{p}+\left\| \mathbf{\Theta } \right\| ^2\sigma ^2\leqslant P_{\mathrm{aRIS}}.
	\end{align}
	Similarly, the subproblem of $\mathbf{p}$ is formulated as
	\begin{subequations}\label{MRM_problem_3}
		\begin{align}
			\min_{\mathbf{p},\varepsilon} \quad& \varepsilon 
			\\
			\mbox{s.t.}\quad& \widehat{R}_k\left( \mathbf{p} \right)\geqslant \frac{l_k{f_{\mathrm{E},k}}}{\varepsilon{f_{\mathrm{E},k}} -l_kc_k},\;\forall k\in \mathcal{K},
			\\
			&\eqref{constraint_p},\eqref{constraint_P_p}.\notag
		\end{align}
	\end{subequations}
	Similar to Problem \eqref{MRM_problem_2}, Problem \eqref{MRM_problem_3} is also a second-order cone programming problem, whose globally optimal solution can also be obtained by existing optimization tools.

\begin{algorithm}[t]
		\caption{BCD algorithm for solving Problem \eqref{MCL_problem}}
	\label{BCD-SCA}
	\textbf{Initialize}: Initialize $\mathbf{F}^1,\boldsymbol{\theta }^1,\mathbf{p}^1,\boldsymbol{l}^1,\boldsymbol{f}_{\mathrm{E}}^1$ to feasible values. Set $n=1$, the maximum number of iterations $n_{\max}$ and the error tolerance  $\zeta$.
	\begin{algorithmic}[1]
		\While{$n\leqslant n_{\max}$ and $\mathcal{T}\left( \mathbf{F}^{n},\boldsymbol{\theta}^{n},\mathbf{p}^{n},\boldsymbol{l}^{n},\boldsymbol{f}_{\mathrm{E}}^{n} \right) \zeta \leqslant$
			{$\left| \mathcal{T}\!\left(\mathbf{F}^{n+1},\boldsymbol{\theta }^{n+1},\mathbf{p}^{n+1},\boldsymbol{l}^{n+1},\boldsymbol{f}_{\mathrm{E}}^{n+1} \right) \!\!-\!\!\mathcal{T}\!\left( \mathbf{F}^{n},\boldsymbol{\theta }^{n},\mathbf{p}^{n},\boldsymbol{l}^{n},\boldsymbol{f}_{\mathrm{E}}^{n} \right) \right|$}}
		\State Calculate $\boldsymbol{l}^{n+1}$ by using \eqref{l_opt};
		\State Calculate $\boldsymbol{f}_{\mathrm{E}}^{n+1}$ by using Algorithm \ref{SCA};
		\State Calculate $\mathbf{F}^{n+1}$ by solving Problem \eqref{f_problem};
		\State Calculate $\boldsymbol{\theta }^{n+1}$ by solving Problem \eqref{MRM_problem_2};
		\State Calculate $\mathbf{p}^{n+1}$ by solving Problem \eqref{MRM_problem_3};
		\State Set $n \gets n+1$;
		\EndWhile
	\end{algorithmic}
\end{algorithm}

\begin{table}[t]\footnotesize
	\vspace{-0.3cm}
		\centering
		\caption{Default Simulation Parameter Setting}
		\label{table_parameter}
		\begin{tabular}{|c|c|}
			\hline
			&  \\ [-6pt]
			Description & Parameter and Value \\
			\hline
			& \\ [-6pt]
			Computing model &\makecell[r]{$[L_1,L_2,L_3]=[250,300,350]$ Kb \\ $[f_{\mathrm{L},1},f_{\mathrm{L},2},f_{\mathrm{L},3}]=[4,5,6]\times10^8$ cycle/s \\ $[c_1,c_2,c_3]=[700,750,800]$ cycle/s \\ $f_{\mathrm{E}}^{\mathrm{tot}}=50\times10^9$ cycle/s}  \\
			\hline
			& \\ [-6pt]
			\makecell[c]{Signal transmission \\ model} &\makecell[r]{$B=1$ MHz, $M=16$, $N=4$\\$\alpha_{\mathrm{UR}}=\alpha_{\mathrm{RA}}=2.2$, $\alpha_{\mathrm{UA}}=2.8$ \\ $\xi=0.8$, $P_{\mathrm{DC}} = -5$ dBm, $P_{\mathrm{c}} = -10$ dBm \\ $\delta ^2=-80$ dBm, $\sigma ^2 = -70$ dBm} \\
			\hline
			& \\ [-6pt]
			Others & \makecell[r]{$n_{\max}=100$, $\zeta=10^{-4}$} \\
			\hline
		\end{tabular}
		\vspace{-0.3cm}
\end{table}

\vspace{-0.5cm}
\subsection{Algorithm Development}
Based on the above discussions, a BCD-based algorithm for solving Problem \eqref{MCL_problem} is proposed, the detail of which are summarized in
Algorithm \ref{BCD-SCA}.
Since the value of the objective function $\mathcal{T}\left(\mathbf{F},\boldsymbol{\theta },\mathbf{p},\boldsymbol{l},\boldsymbol{f}_{\mathrm{E}}\right) $ is nondecreasing
at each iteration and has an upper bound, the convergence of Algorithm \ref{BCD-SCA} is guaranteed.
The computational complexity mainly depends on solving the subproblems  \eqref{f_problem} and \eqref{MRM_problem_2}.
According to \cite{ben2001lectures}, the computational complexity of Problem \eqref{f_problem} is $\mathcal{O}\left( N^3 \right)$.
In addition, the complexity of solving Problem \eqref{MRM_problem_2} is given by $\mathcal{O}\left( MK^{3.5}+M^3K^{2.5} \right) $.
Denote the number of iterations required for Algorithm 1 to converge by $r_{\mathrm{act}}$.
Hence, the overall computational complexity of Algorithm 2 for each iteration is $\mathcal{O}\left( r_{\mathrm{act}}N^3+MK^{3.5}+M^3K^{2.5} \right) $.

\vspace{-0.4cm}
\section{Simulation Results}\label{Simulation Results}

In the simulation scenario, the locations of  the RIS and the AP are set at $($260 m, 0$)$ and $($0, 0$)$, respectively.
	Three users with computation offloading power $p_k=1$ mW are randomly located in a 10 m $\times$ 10 m area centered at $($280 m, 10 m$)$.
	The heights of the users, the RIS and the AP are set to 1.5 m, 10 m and 30 m, respectively.
	The path loss exponents of the reflected user-RIS, RIS-AP and user-AP channel are set to $\alpha_{\mathrm{UR}}$, $ \alpha_{\mathrm{RA}}$ and $\alpha_{\mathrm{UA}}$, respectively.
	For $q\in\left\lbrace \mathrm{UR}, \mathrm{RA}, \mathrm{UA}\right\rbrace $, the large-scale path loss is modelled as $\mathrm{PL}_q=-10\alpha_q \log _{10}d_q -30 $ dB, where $d_q$ is the distance of link $q$.
	The default settings of these parameters are specified in
	the “Signal transmission model” block of Table \ref{table_parameter}.
	Besides, the variables $L_k$, $f_{\mathrm{L},k}$, $c_k$ and $f_{\mathrm{E}}^{\mathrm{tot}}$ are specified in the “Computing model” block of Table \ref{table_parameter}.
	The default setting of maximum number of iterations $n_{\max}$ and error tolerance $\zeta$ are specified in the “Others” block of Table \ref{table_parameter}.
	We consider practical scenarios, where 2-bit control of each reflecting element is used in our simulation, where each optimal phase shift $\phi_m^{\mathrm{opt}}$ generated by Algorithm \ref{BCD-SCA} is quantized as 0, $\pi/2$, $\pi$ or $3\pi/2$.
	In addition, we investigate the passive RIS as a performance benchmark, where the amplification power budget is modelled as $P_{\mathrm{pRIS}}=P_{\mathrm{tot}}-MP_{\mathrm{c}}$ \cite{9377648}, and the subproblem for optimizing the passive reflection coefficients is solved by using the popular semidefinite relaxation method.

\begin{figure}
	\vspace{-1.2cm}
	\centering
	\includegraphics[width=0.8\linewidth]{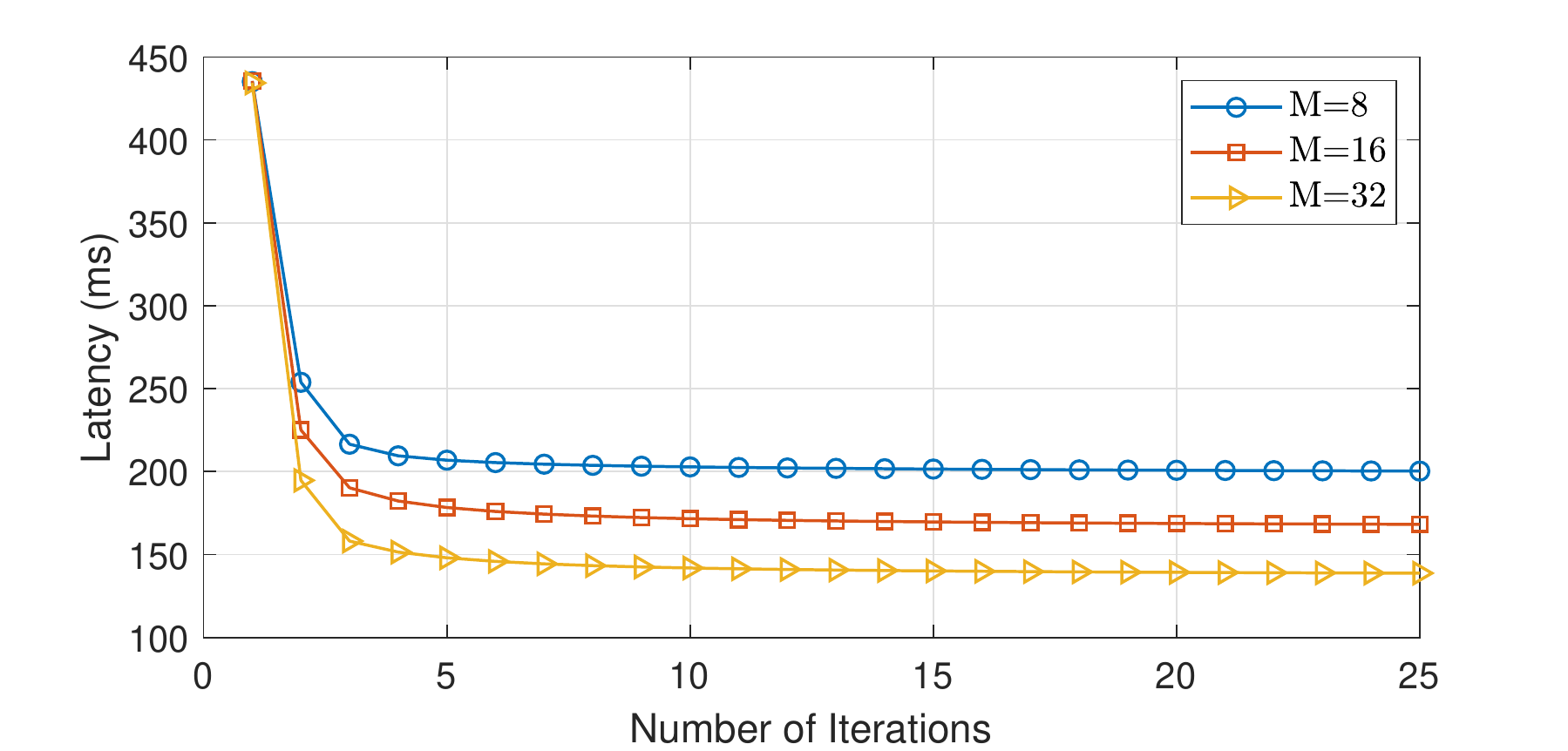}
	\vspace{-0.3cm}
	\caption{Convergence of the proposed algorithm for $M=[8,16,32]$.}
	\label{iteration_fig}
	\vspace{-0.45cm}
\end{figure}

Fig. \refeq{iteration_fig} depicts the convergence behaviour of Algorithm \ref{BCD-SCA} by fixing the amplification power budget to $P_{\mathrm{aRIS}} = 10$ mW.
It can be observed that although the convergence will be slightly slower as the number of reflection coefficients increases, the algorithm can converge within 15 iterations, which demonstrates the effectiveness of the proposed algorithm.

\begin{figure}
	\centering
	\includegraphics[width=0.8\linewidth]{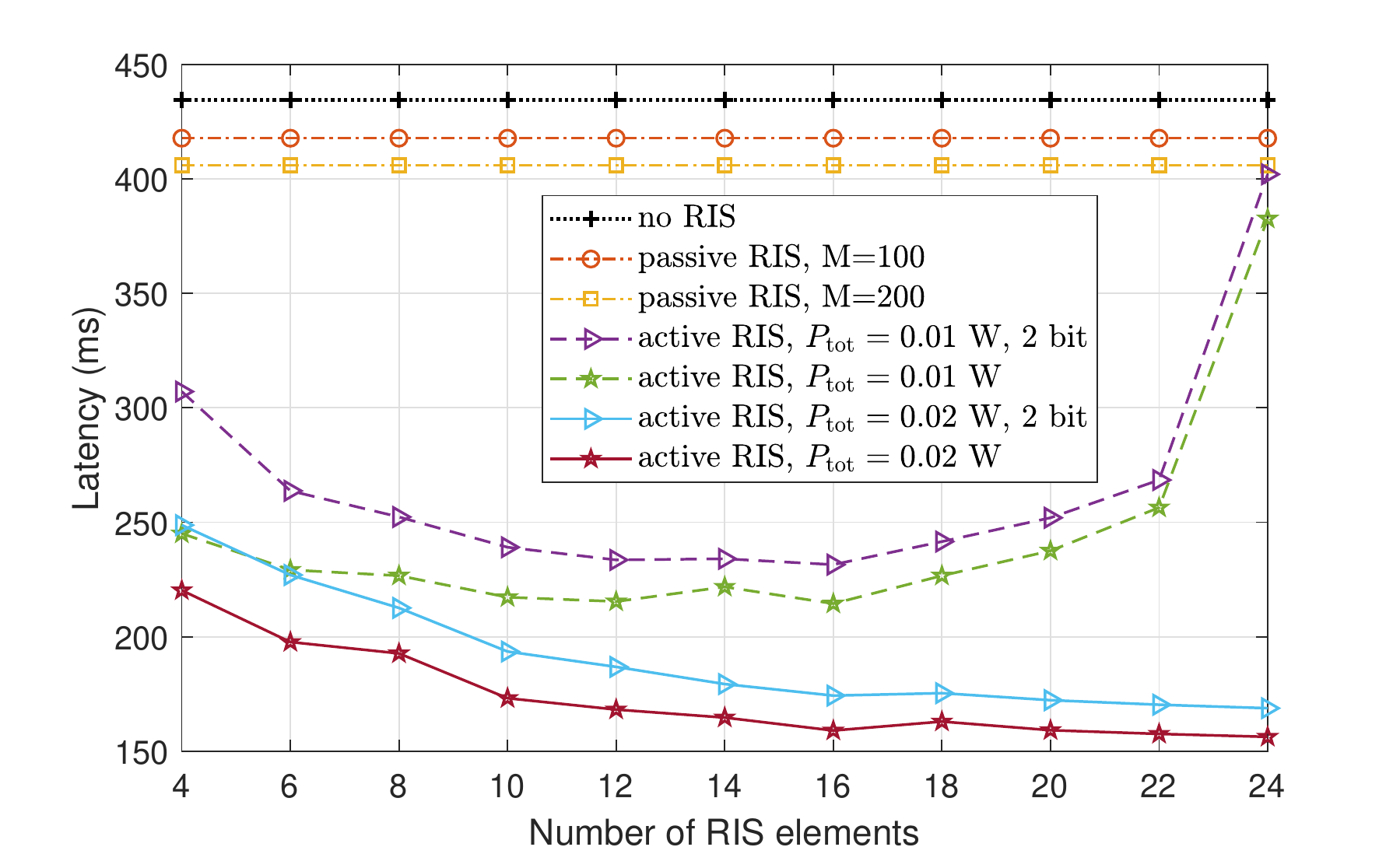}
	\caption{Impact of RIS reflecting element number.}
	\label{M_fig}
	\vspace{-0.7cm}
\end{figure}
\begin{figure}
	\vspace{-1.4cm}
	\centering
	\includegraphics[width=0.8\linewidth]{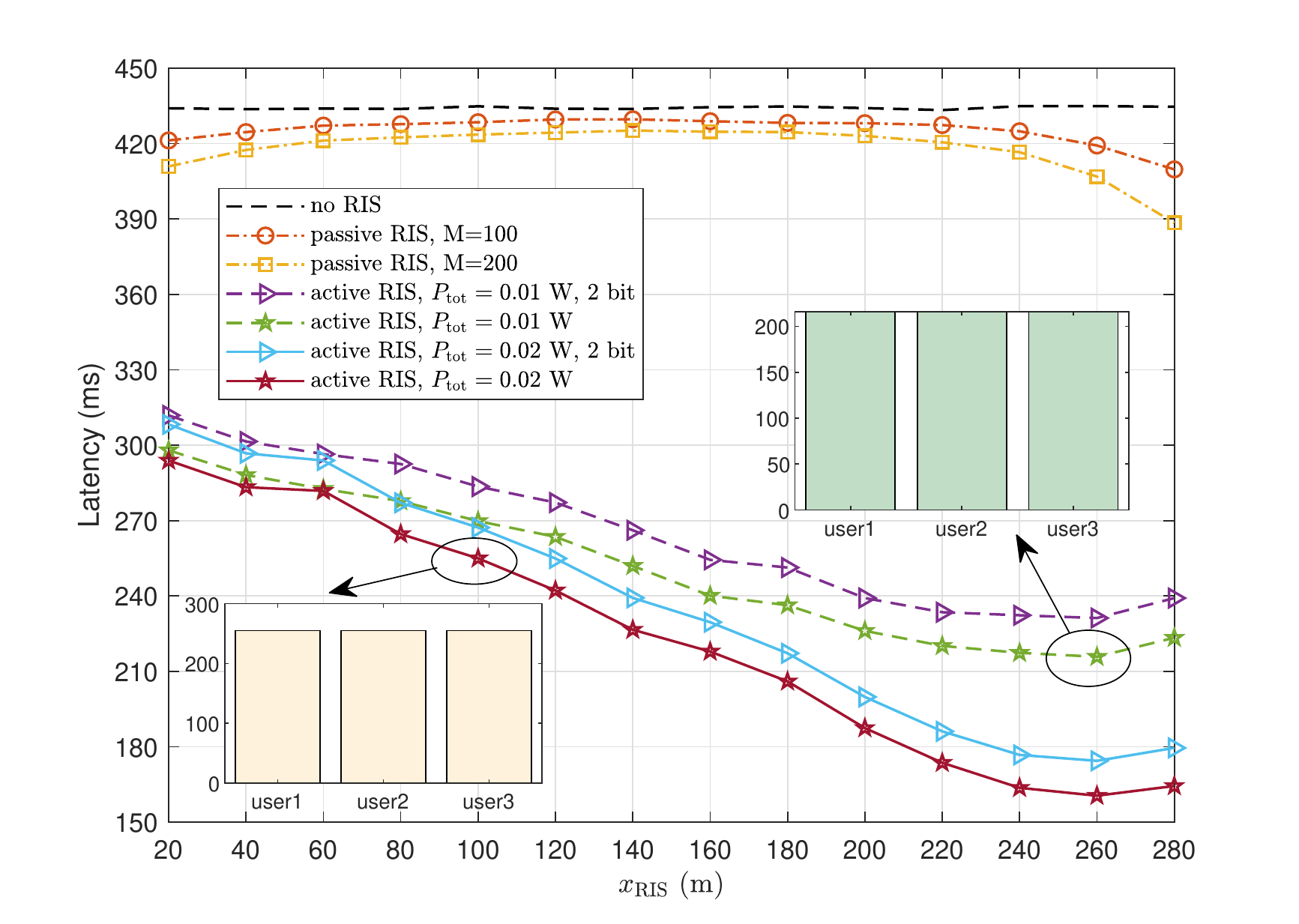}
    \vspace{-0.4cm}
	\caption{Impact of the RIS location for $P_{\mathrm{tot}}=[10, 20]$ mW.}
	\label{location_fig}
	\vspace{-0.8cm}
\end{figure}

Fig. \ref{M_fig} and Fig. \ref{location_fig} plot the latency versus the number of reflecting elements and the x coordinate of the RIS $x_{\mathrm{RIS}}$, respectively.
The total power budget is set to $P_{\mathrm{tot}}=$ 10 and 20 mW.
In these cases, the maximum numbers of reflecting elements the passive RIS can employ are 100 and 200, respectively.
From Fig. \ref{M_fig}, it is observed that when the direct link is not severely obscured, compared to the schemes with passive RIS, the schemes with active RIS and the corresponding {2 bit} schemes can reduce the required number of reflecting elements while achieving lower latency.
Hence, active RIS is expected to effectively overcome the double path loss attenuation, and offers advantages in terms of size and deployment flexibility.
The results in Fig. \ref{location_fig} again demonstrate the advantages of active RIS in improving the performance of MEC systems.
Moreover, it is shown that the achieved latency increases as the distance between the active RIS and the users increases.
This is because the noise power $\sigma ^2$ at the active RIS is assumed to be constant, while the incident signal power is stronger when the RIS is located near the users.
Therefore, the best place to deploy active RIS is close to the users in the MEC system.
In addition, Fig. \ref{location_fig} illustrates the individual latency, which shows the fairness achieved by our proposed algorithm.

\vspace{-0.5cm}
\section{Conclusion} \label{conclusion}
In this work, we proposed to reduce the MCL of the MEC system by deploying an active RIS.
A joint computing and communication design was proposed to solve the MCL minimization problem, subject to both the active RIS phase shift constraints and the edge computing capability constraints.
The simulation results showed that under the condition of the same power budget, the active RIS can achieve much better performance than the passive RIS.

%
\vspace{-0.5cm}
\bibliographystyle{IEEEtran}
\bibliography{IEEEabrv,Refer}

\end{document}